# The Effect of Epsilon-Near-Zero (ENZ) Modes on the Casimir Interaction between Ultrathin Films

Tao Gong,[1,2] Iñigo Liberal,[3] Benjamin Spreng,[1] Miguel Camacho,[4] Nader Engheta,[5,*] and Jeremy N. Munday[1,*]

[1]*Department of Electrical and Computer Engineering, University of California, Davis, USA*
[2]*Department of Materials Science and Engineering, University of California, Davis, USA*
[3]*Institute of Smart Cities, Public University of Navarre, Spain*
[4]*Department of Electronics and Electromagnetism, University of Seville, Spain*
[5]*Department of Electrical and Systems Engineering, University of Pennsylvania, USA*

**Abstract**

Vacuum fluctuation-induced interactions between macroscopic metallic objects result in an attractive force between them, a phenomenon known as the Casimir effect. This force is the result of both plasmonic and photonic modes. For very thin films, field penetration through the films will modify the allowed modes. Here, we investigate the Casimir interaction between two ultrathin films from the perspective of the force distribution over real frequencies for the first time and find pronounced repulsive contributions to the force due to the highly confined and nearly dispersion-free epsilon-near-zero (ENZ) modes that only exist in ultrathin films. These contributions are found to persistently occur around the ENZ frequency of the film and are irrespective of the inter-film separation. We further associate the ENZ modes with a striking thickness dependence in the averaged force density for conductive thin films, a metric signifying a thin-film's acceleration due to Casimir effect. Our results shed light on the role of the unique vacuum fluctuation modes existing in ultrathin ENZ materials, which may offer significant potential for engineering the motion of objects in nanomechanical systems.

When two charge-neutral plates are brought close to each other to form a cavity, a force arises due to quantum vacuum fluctuations of electromagnetic fields between the plates, as a result of the spatial mismatch of the energy between the allowed modes inside the cavity and those outside of it. This force was first predicted by Casimir in 1948 [1], and significant advances have been achieved in both theoretical prediction and experimental measurement of Casimir forces with different configurations over the past decades [2-5]. In parallel, Casimir interactions have been exploited in a plethora of scenarios such as quantum mechanical actuation [6], parametric amplification [7], quantum levitation/trapping [8], dissipation dilution [9], and macroscopic self-assembly processes [10]. Particularly, the combinatorial influence of the constituent materials and the structural geometries on the force is of fundamental and technological interests, as it can possibly lead to Casimir repulsion or non-standard dependence of the force on inter-object separations [11-16].

Amongst various geometries, ultrathin films are surprisingly rarely studied. For ultrathin films, Casimir interactions take place within a very small material volume, with important electromagnetic and mechanical implications. An intuitive way of thinking about how the interaction is modified between two ultrathin metallic films compared with bulk counterparts is through the skin-depth effect [2,4]: if the film thickness is less than the skin-depth of the metal (about 10 nm for most metals using visible light), the plates are much less reflective. According to the Lifshitz theory, the Casimir force scales with the plate reflectivity in the wavelength range from UV to far-IR at sub-micron inter-plate separations [17]. Consequently, the force is expected to be notably smaller than the one between bulk reflective metals. The skin-depth effect in Casimir force has indeed been confirmed experimentally with palladium thin films on transparent substrates [18,19]. However, as the film thickness is reduced to the deeply subwavelength range (typically a few nanometers), a new polaritonic "epsilon-near-zero (ENZ) mode" emerges around a particular frequency with vanishing permittivity, i.e. the ENZ frequency [20-22]. This mode has several unique properties including nearly zero-dispersion and high field confinement. Yet to our knowledge, the effect of this important mode on the Casimir force has hitherto not been determined.

In this work, we examine the Casimir interactions between two symmetric ultra-thin films from the perspective of the force spectral distributions over real frequencies. We reveal the effect of the ENZ modes on the Casimir force, to the best of our knowledge for the first time, showing that there is a significant repulsive contribution pinned around the ENZ frequency caused by the ENZ modes, despite the net force always being attractive. Additionally, we define an alternative figure of merit (FOM) to the total force for the system: the averaged Casimir force per unit plate volume exerted on a thin film (i.e., the averaged force density). This FOM is proportional to the thin-film's acceleration due to the Casimir force and hence designates important nanomechanical characteristics. We find that the variation of the FOM with the film thickness hinges critically on the optical properties of the composing material. Surprisingly, the FOM increases monotonically with reduced film thickness down to sub-nanometer range for a conductive film whereas an optimal thickness exists for an insulating film to achieve a maximal FOM. This distinction is ascribed to the ENZ modes as well as the accompanying surface plasmon modes in ultrathin conductive films. Our results provide new insight on the underlying relation between Casimir interactions and the special optical modes in ultrathin ENZ films, which will bring about an alternative perspective of controlling the motion of nano-scale objects in nanomechanical systems.

To relate the Casimir forces to the optical modes between two objects, force calculation with respect to real frequencies is necessary. Traditionally, the well-known Lifshitz theory used for force calculation between two parallel surfaces associates the force with the reflection coefficients of the surfaces in complex frequency through the Wick rotation, $\omega = i\xi$. While the computation using complex frequency is mathematically easier because of the monotonical decay of the dielectric function and hence of the force integrand with $\xi$, the physical interpretation of the role played by the individual optical modes is quite elusive [23]. Instead, we determine the spectral distributions of the Casimir force over real frequencies (also referred to as force spectra in this paper) between two identical thin films by applying the Lifshitz equation in terms of real frequency and momentum and then integrating over all momenta for a given frequency [24,25], allowing us to possibly resolve the contributions from different modes. Figure 1(a) and 1(b) show the force spectra for a film thickness of $t = 200$ nm (optically thick film) and $t = 2$ nm (ultrathin film), respectively, at a fixed inter-plate separation $d = 10$ nm. Each film is made of a conductive material with a Drude model as its dielectric function: $\varepsilon(\omega) = 1 - \frac{\omega_p^2}{\omega^2 + i\gamma_p \omega}$, where $\omega_p = 3$ eV and $\gamma_p = 0.035$ eV are the plasma and damping frequencies, respectively.

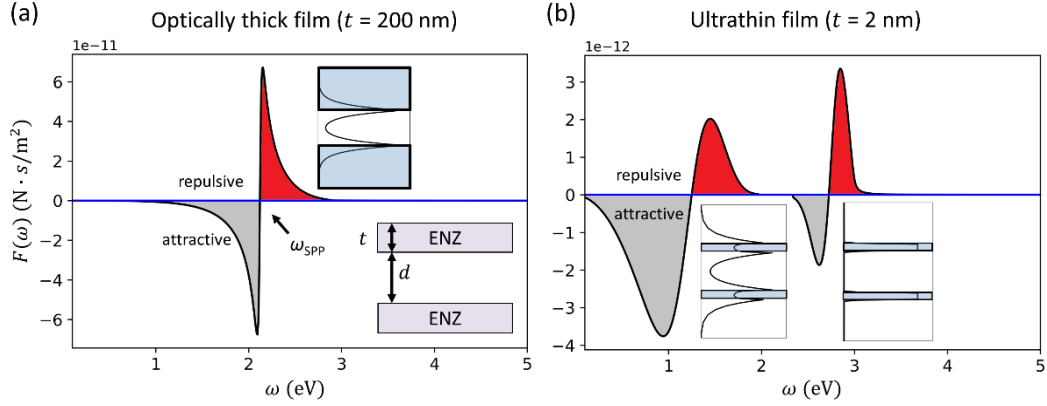

FIG. 1. Spectral distributions of the Casimir pressure between two symmetric films with film thicknesses of (a) $t = 200$ nm (optically thick) and (b) $t = 2$ nm (ultrathin). The inter-plate separation is $d = 10$ nm. The Drude model parameters are: $\omega_p = 3$ eV and $\gamma_p = 0.035$ eV. For the optically thick films, the force spectrum features a single resonant peak-valley pair around the SPP resonance frequency $\omega_{SPP} \sim 2.12$ eV due to the coupled SPP modes. Inset in (a) is a schematic of the mode field intensity profile. While for the ultrathin films, two resonant peak-valley pairs emerge due to the coupled ENZ modes (at ~3 eV) and the short-ranged SPP modes at a lower frequency (~1.25 eV). Inset of (b) shows schematics of field intensity profiles for the two modes (right for the coupled ENZ modes and left for the coupled short-ranged SPP modes, respectively).

The force spectral distribution exhibits a distinctive difference between optically thick and ultrathin films. For the former (Fig. 1(a)), the force spectrum features a single resonant peak-valley pair resulting from the Coulomb interaction between the surface plasmon polaritons (SPP) at the inner surface of the two plates [26]. The interaction splits the two otherwise standalone SPPs into two coupled (hybrid) modes: anti-symmetric (anti-binding) SPP mode with a higher energy, and symmetric (binding) SPP mode with a lower energy [27,28]. The inter-plate separation controls the strength of the interaction between the two standalone SPPs and determines both the amplitude of the force spectral resonance and where the resonant peak-valley pair is located on the frequency axis. This is because the force spectral resonance occurs where the local density of electromagnetic states (and the associated energy) is altered the most compared with the case when the two plates separated infinitely far apart [27]. As shown in Fig. 1(a), with a 10 nm separation the resonance in the force spectrum occurs around the SPP resonance frequency $\omega_{SPP} = \frac{\omega_p}{\sqrt{2}} \sim 2.12$ eV. Larger separation (i.e., weaker SPP coupling) results in the frequency of the resonant force peak-valley pair being further from the undisturbed SPP resonance frequency $\omega_{SPP}$ at a single film surface (as represented by Fig. S1(a)-S1(c)) [28]. Additionally, the anti-symmetric SPP mode (higher energy) consistently yields a peak (repulsion), while the symmetric SPP mode (lower energy) always yields a valley (attraction). The inset of Fig. 1(a) shows a schematic of the field intensity profile of the coupled SPP modes.

The force spectral distribution for ultrathin films exhibits two resonant peak-valley pairs rather than one as is the case for optically thick films (see Fig. 1(b)). We attribute this behavior to the emergence of a unique mode in ultrathin films, the ENZ mode. It arises when the film thickness $t$ is smaller than 1/50 of the characteristic plasma wavelength: $\lambda_p = \frac{2\pi c}{\omega_p}$ [22]. The ENZ mode stems from the strong coupling of the otherwise unperturbed SPPs on both interfaces of the isolated film. The mode frequency is almost pinned at the ENZ frequency, with other intriguing properties such as extremely high field confinement and near free-of-dispersion [22,29]. When two ultrathin films are brought in proximity of each other, the ENZ modes in the opposing films interact with each other, resulting in a resonant peak-valley pair around the ENZ frequency (~3 eV) in the force spectrum. Concurrently, the ENZ mode in an ultrathin film is always accompanied by the presence of a highly dispersive mode called the short-ranged SPP mode at a lower frequency (~1.25 eV for our case) [20]. As a result, another resonant peak-valley pair in the force spectrum is observed due to the coupled short-ranged SPPs when the two plates are brought close. The schematics of the coupled ENZ modes (right) and the short-ranged SPP modes (left) are shown in the inset of Fig. 1(b).

The coupled ENZ modes have two prominent distinctions when compared to other modes. Firstly, the ENZ modes-induced resonance is persistently pinned around the ENZ frequency, almost invariant with the inter-plate separation, $d$ (Fig. 2a). On the contrary, the short-ranged SPP modes-induced resonance gradually redshifts with increasing separation, akin to the regular SPP modes-induced one for thick films (also see Fig. S1(d)-Fig. S1(f)). Secondly, the overall contribution by the ENZ modes to the total force, which is calculated by the integral of the ENZ-mode force spectrum over the frequency range that spans the ENZ-mode-induced resonance peak-valley pair, is consistently repulsive regardless of the inter-plate separation, whereas the short-ranged SPPs (or long-ranged SPPs for thick films) lead to attraction at all separations. In addition, while the overall force is always attractive between two symmetric objects [30], the relative magnitude of the contributions due to the two respective modes (ENZ and SPP) do vary with separation. Figure 2(b) reveals that the magnitude of the ENZ modes contribution is smaller than the short-ranged SPP modes when the inter-plate separation is small, yielding an overall attractive force. For larger inter-plate separations, the repulsive ENZ modes outweighs the short-ranged attractive SPP modes; however, the total force is still attractive due to the large overall attractive contributions by propagating waves (such as the waveguide modes allowed between the two plates separated sufficiently far away) [28].

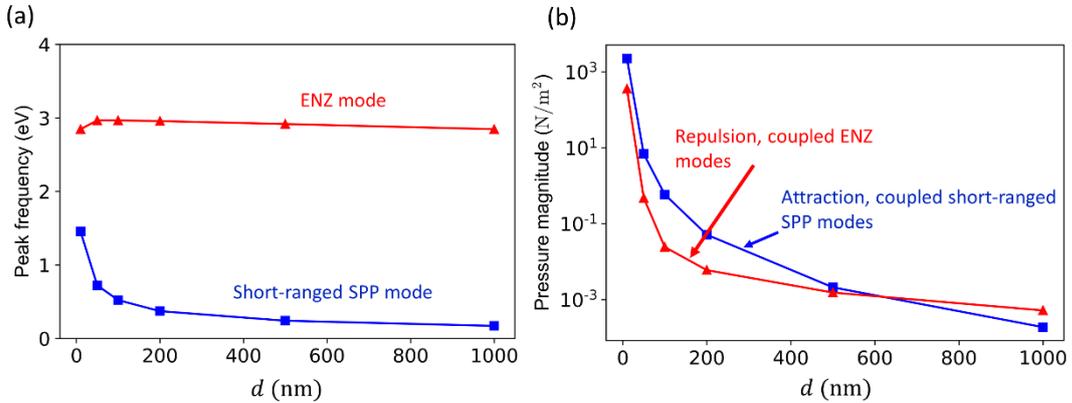

FIG. 2. Contributions to the Casimir force between two ultrathin films by the coupled ENZ modes and the short-ranged SPP modes. The film thickness $t = 2$ nm. (a) Corresponding frequencies to the force spectral peak in the resonant peak-valley pair by each mode, and (b) The overall contribution to the force from each mode. The coupled ENZ modes (red) in the vicinity of the ENZ frequency always yield repulsion irrespective of the inter-plate separation, while the coupled short-ranged SPP modes (blue) always give rise to attraction.

Because of the presence of the ENZ and the SPP modes in conductive films, the film thickness effect on the Casimir force is fundamentally different for insulating films. As shown in Fig. 3(a), what both types of films have in common is that the thicker the films are, the larger the overall force is. This can be intuitively explained by comparison to an optical cavity where fewer photons leak out of the cavity when the film thickness is larger. At the same time, conductive films exhibit a larger force in all studied configurations, and the difference between conductive and insulating films increases as the thickness decreases. Note that the dielectric function for the conductor uses the same Drude model as defined above, while that for the insulator is modeled using a single Lorentzian oscillator as $\varepsilon(\omega) = 1 + \frac{C_L \omega_L^2}{\omega_L^2 - \omega^2 - i\gamma_L \omega}$ with the parameters $C_L = 1$, $\omega_L = 15$ eV and $\gamma_L = 0.01$ eV. However, because the electromagnetic field is not evenly distributed inside the film, one can expect that as the film thickness varies, the force magnitude may not scale on equal footing for conductive and insulating films. To better depict this behavior, we introduce a figure-of-merit (FOM) that consists of the averaged force volume density, defined as the averaged Casimir pressure per unit film thickness, or equivalently the averaged Casimir force per unit volume of a thin film. The proposed FOM translates directly to the acceleration of the thin-film object if we let it be free under the influence of the Casimir force, provided the mass density is uniform across the film. Therefore, the FOM has a clear physical connection to the mechanical response of the system to the Casimir force.

Figure 3(b) shows how the FOM scales with the film thickness at several inter-plate separations. The FOM for a conductive film (black solid curves) monotonically decreases with increasing thickness. This behavior is due to the fact that in a conductive film, the electromagnetic fields are confined at the surface and decay exponentially away from it (i.e. it can be approximated as $I(x) \sim e^{-\alpha x}$, where $x$ denotes the distance from the surface inside the film and $\alpha$ denotes the field attenuation coefficient). Therefore with reduced film thickness, the fields are increasingly concentrated per unit volume. At a small separation (e.g., $d = 10$ nm), the FOM can be approximated as $\propto \int_0^t e^{-\alpha_{\text{SPP}} x} dx / t = (1 - \exp(-\alpha_{\text{SPP}} t))/\alpha_{\text{SPP}} t$. When the film is optically thick ($t \gg \alpha_{\text{SPP}}^{-1}$), FOM $\propto 1/\alpha_{\text{SPP}} t$ (see the right dashed line in Fig. 3(c)). By contrast, in the ultrathin limit, the emergence of the ENZ mode flattens the field distribution across the film for that mode ($t \ll \alpha_{\text{ENZ}}^{-1}$), while the short-ranged SPP still results in a notable exponential field decay inside the film ($t > \alpha_{\text{SR-SPP}}^{-1}$). This additional ENZ mode therefore lowers the slope of the FOM with respect to $t$ in a logarithmic scale representation (see the left dashed line in Fig. 3(c)).

By contrast, we find that the FOM for an insulating film maximizes at a particular thickness. The optimal thickness $t_{\text{opt}}$ as a function of inter-plate separation $d$ is shown in Fig. 3(d), where a nearly linear relation is revealed. A physically intuitive interpretation can be obtained as follows. In this case, there is no surface (evanescent) mode existing in the relevant frequency range; instead, the fluctuating electromagnetic fields (sometimes referred to as virtual photons) can transmit into the film and are subject to optical interference due to multiple reflections at both interfaces, resulting in a non-monotonic spatial variation of the fields across the film. The largest contribution to the force occurs at the frequency $\omega_{\text{ph}} \sim \frac{1}{d}$ [31]. Therefore, the field intensity inside the film is maximized at the location $x \propto \frac{\lambda_{\text{ph}}}{n} = \frac{2\pi c}{n \omega_{\text{ph}}} \propto \frac{d}{n}$, where $n$ is the refractive index of the insulating material. As such, the FOM maximizes with a thickness $t_{\text{opt}} \propto d/n$. For practical considerations, we show the calculated FOMs for Au and $SiO_2$ using realistic optical parameters in Fig. S2. They are in excellent agreement with the behaviors obtained using Drude and Lorentz model, respectively.

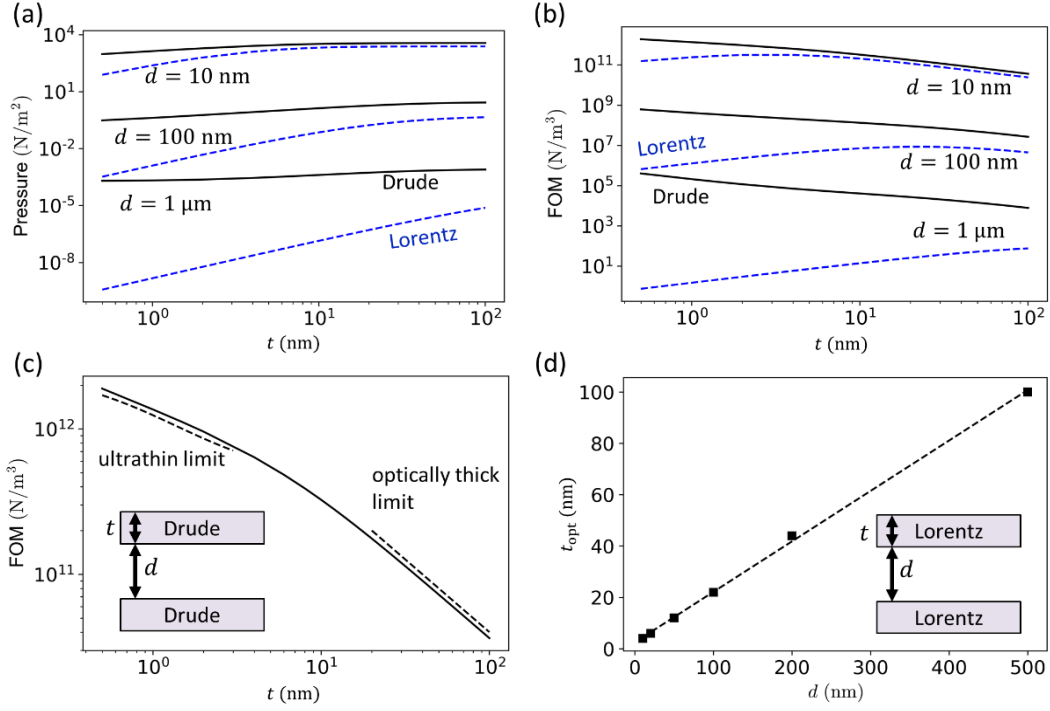

FIG. 3. Film thickness effect on the Casimir interaction for conductive and insulating films. (a) Casimir pressure and (b) the force density FOM as a function of film thickness $t$ for different inter-plate separations of $d = 10$ nm, $d = 100$ nm, and $d = 1$ μm. The conductive (Drude model) and insulating (Lorentz model) films are represented by black solid and blue dashed curves, respectively. (c) The FOM scaling with $t$ for conductive films at $d = 10$ nm, with asymptotes (dashed lines) for the ultrathin limit where ENZ modes arise and for the optically thick limit. (d) The optimal thickness $t_{\text{opt}}$ yielding the maximum FOM as a function of inter-plate separation $d$ for the insulating material.

We note that the scaling of the FOM for a conductive film with the film thickness is distinct from that of many other physical phenomena, e.g., the scattering cross section of a nanoparticle generally decreases with shrinking object sizes. This peculiar scaling behavior indicates that if we released two conductive thin films in a parallel plate configuration, thinner films would be more prone to move towards each other, while for dielectric films there is a "sweet spot" for maximizing the effect, which could lead to finer control and engineering of the propelling motion of thin-film objects via Casmir actuation (or from the opposite perspective, could help to design systems that combat the stiction of movable parts in nanomechanical systems). One might also expect faster harmonic oscillations for an ultrathin conductive film than for an optically thick one in a quantum-levitating system driven by the Casimir effect when the object is perturbed from its stable equilibrium.

One caveat is that other effects could become more influential when the film thickness is down to the level of a few nanometers or smaller. For example, quantum confinement discretizes the electron energy bands, which can cause non-negligible changes and bring in anisotropy of the optical property of the thin film [32-35]. In addition, conductive films can become discontinuous in morphology with extremely small thickness, and it has been reported that gold films can turn into insulating state below the percolation threshold (typically a few nanometers) [36-38]. Consequently, all of these factors would need to be taken into account for the design of MEMS/NEMS components.

In conclusion, we demonstrate how the special ENZ modes supported by ultrathin films contribute to the Casimir force. We find that a significant fraction of the contribution to the force originates from the highly confined, near-dispersion-less ENZ mode, which is pinned at the ENZ frequency regardless of the plate-plate separation. This mode persistently yields a repulsive interaction, despite the total force being attractive. As a result of the ENZ mode and the other SPP modes, the scaling of the Casimir force with film thickness between conductive films behaves radically differently from the force between insulating films, with the monotonically increasing average force density for the former and non-monotonic scaling for the latter. Our results provide new insights on the Casimir interaction in ultrathin films and will help to guide the design and control of nanoscale thin-film components for nanomechanical systems.

**Acknowledgement**

The authors acknowledge financial support from the Defense Advanced Research Program Agency (DARPA) QUEST program No. HR00112090084.

* engheta@ee.upenn.edu

* jnmunday@ucdavis.edu

# Supplementary Figures for: The Effect of Epsilon-Near-Zero (ENZ) Modes on the Casimir Interaction between Ultrathin Films

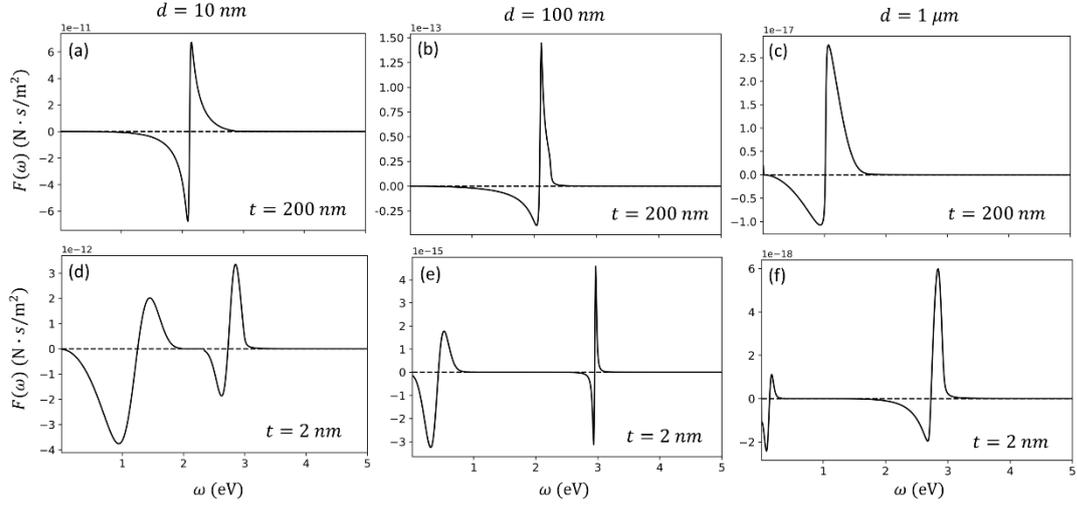

FIG. S1. Spectral distributions of the Casimir pressure due to surface modes between (a-c) optically thick films with the film thickness $t = 200$ nm, and (d-f) ultrathin films with the film thickness $t = 2$ nm, at inter-plate separations of $d = 10$ nm, $d = 100$ nm and $d = 1$ μm. A significant force spectral resonance is consistently observed around the ENZ frequency due to the coupled ENZ modes for the ultrathin films, while the resonance due to coupled SPP modes for the 200 nm-thick films occurs at varying frequencies as the inter-plate separation changes as a result of the variation of the coupling strength.

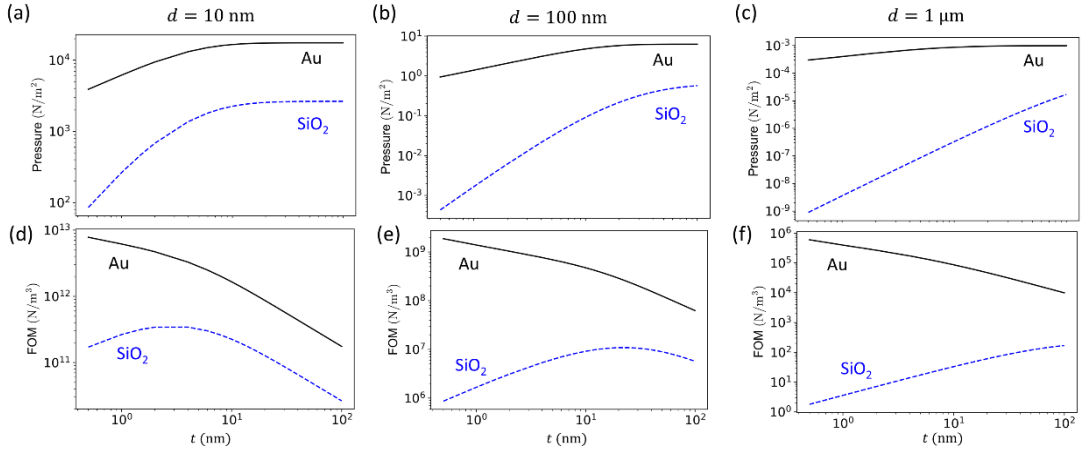

FIG. S2. (a-c) Casimir pressure and (d-f) the derived FOM as a function of film thickness $t$ for at inter-plate separations of $d = 10$ nm, $d = 100$ nm and $d = 1$ μm for Au and SiO$_2$.